\documentclass[11pt,a4paper,useAMS,usenatbib]{emulateapj}
\usepackage{epsfig}
\usepackage{amsmath}
\usepackage{natbib}

\begin{document}

\title{Exploring the unusually high black hole-to-bulge mass ratios in NGC4342 and NGC4291: \\  the asynchronous growth of bulges and black holes}

\author{\'Akos Bogd\'an\altaffilmark{1}, William R. Forman\altaffilmark{1}, Irina Zhuravleva\altaffilmark{2}, J. Christopher Mihos\altaffilmark{3}, Ralph P. Kraft\altaffilmark{1}, Paul Harding\altaffilmark{3}, Qi Guo\altaffilmark{4,5}, Zhiyuan Li\altaffilmark{1}, Eugene Churazov\altaffilmark{2}, Alexey Vikhlinin\altaffilmark{1}, Paul E. J. Nulsen\altaffilmark{1}, Sabine Schindler\altaffilmark{6}, and Christine Jones\altaffilmark{1}}

\affil{\altaffilmark{1}Smithsonian Astrophysical Observatory, 60 Garden Street, Cambridge, MA 02138, USA}
\affil{\altaffilmark{2}Max-Planck-Institut f\"ur Astrophysik, Karl-Schwarzschild-str. 1, 85741 Garching bei M\"unchen, Germany}
\affil{\altaffilmark{3}Department of Astronomy, Case Western Reserve University, 10900 Euclid Avenue, Cleveland, OH 44106, USA}
\affil{\altaffilmark{4}Partner Group of the Max Planck Institute for Astrophysics, National Astronomical Observatories, Chinese Academy of Sciences, Beijing, 100012, China}
\affil{\altaffilmark{5}Department of Physics, Institute for Computational Cosmology, University of Durham, South Road, Durham DH1 3LE, UK}
\affil{\altaffilmark{6}Institut f\"ur Astro- und Teilchenphysik, Universit\"at Innsbruck, Technikerstrasse 25, 6020 Innsbruck, Austria}
\email{E-mail: abogdan@cfa.harvard.edu}

\shorttitle{NON-CONCURRENT GROWTH OF BULGES AND BLACK HOLES}
\shortauthors{BOGD\'AN ET AL.}

\begin{abstract}
We study two nearby, early-type galaxies, NGC4342 and NGC4291, that host unusually massive black holes relative to their low stellar mass. The observed black hole-to-bulge mass ratios of NGC4342 and NGC4291 are $6.9^{+3.8}_{-2.3}\%$ and $1.9\%\pm0.6\%$,  respectively, which significantly exceed the typical observed ratio of $\sim$$0.2\%$. As a consequence of the  exceedingly large black hole-to-bulge mass ratios, NGC4342 and NGC4291 are $\approx$$5.1\sigma$ and $\approx$$3.4\sigma$ outliers from the $M_{\bullet}-M_{\rm{bulge}}$ scaling relation, respectively. In this paper, we explore the origin of the unusually high black hole-to-bulge mass ratio.  Based on \textit{Chandra} X-ray observations of the hot gas content of NGC4342 and NGC4291, we compute gravitating mass profiles, and conclude that both galaxies reside in massive dark matter halos, which extend well beyond the stellar light. The presence of  dark matter halos around NGC4342 and NGC4291 and a deep optical image of the environment of NGC4342 indicate that tidal stripping, in which $\gtrsim$$90\%$ of the stellar mass was lost, cannot explain the observed high black hole-to-bulge mass ratios. Therefore, we conclude that these galaxies formed with low stellar masses, implying that the bulge and black hole did not grow in tandem. We also find that the black hole mass correlates well with the properties of the dark matter halo, suggesting that dark matter halos may play a major role in regulating the growth of the supermassive black holes. 
\end{abstract}


\section{Introduction}
Observational studies of nearby galaxies have demonstrated a rather tight correlation between the mass of the central supermassive black hole ($M_{\bullet}$) and the mass ($M_{\rm{bulge}}$) \citep{magorrian98,haring04} or central velocity dispersion  ($\sigma_{\rm{c}}$) \citep{gebhardt00,ferrarese00,gultekin09} of the host spheroids. To explain the small intrinsic scatter in the  $M_{\bullet}-M_{\rm{bulge}}$ and  $M_{\bullet}-\sigma_{\rm{c}}$ relations the theoretical paradigm has emerged, in which galaxy bulges and supermassive black holes co-evolve and regulate each other's growth \citep[e.g.][]{silk98,fabian99,kauffmann00,king03,dimatteo05,hopkins06}. However, the masses of supermassive black holes also correlate well with the dark matter halos of galaxies \citep{ferrarese00}, suggesting that it may be the dark matter halo that (indirectly) governs the black hole growth \citep{volonteri11}. Additionally, numerical studies indicate that the $M_{\bullet}-M_{\rm{bulge}}$ relation can be reproduced by non-causal processes, which do not require the co-evolution of black holes and bulges \citep{peng07,jahnke11}. 

\begin{figure*}
 \begin{center}
       \includegraphics[width=8.5cm]{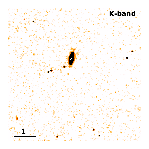}
\hspace{0.2cm}
      \includegraphics[width=8.5cm]{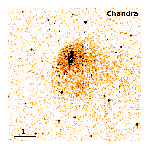}
       \caption{\textit{Left:} 2MASS K-band image of a $6\arcmin \times 6\arcmin$ (or $40 \times 40 $ kpc) region around NGC4342. The center of the galaxy is marked with a white cross. The distribution of stellar light appears to be symmetric, thereby indicating the undisturbed nature of NGC4342. The main optical body is fairly small, $\approx$$90\%$ of the stellar light is concentrated within a $0.7\times2.2$ kpc elliptic region. \textit{Right:} $0.5-2$ keV band merged \textit{Chandra} image of the same region. One pixel corresponds to $\approx$$2$ arcsec. The X-ray emission has a very broad distribution, significantly exceeding that of the stellar light.  The X-ray emission shows a distinct surface brightness edge and a tail, which we identify as a contact discontinuity or cold front -- the nature of the edge is discussed in \citet{bogdan12}. North is up and east is left.}
     \label{fig:ngc4342}
  \end{center}
\end{figure*}

In essence, despite the existence of scaling relations between the black hole mass and the physical properties of host bulges and halos, the origin of these relations remains a matter of debate. In this paper we employ X-ray, optical, and near-infrared observational data of two nearby early-type galaxies with unusually high black hole-to-bulge mass ratios to gain insight in the evolution of their bulges and black holes. 

The main focus of the work is NGC4342, a lenticular galaxy \citep{devaucouleurs91}, which exhibits both an outer and a stellar nuclear disk \citep{bosch98}, and hosts a classical bulge \citep{cretton99,graham01,kormendy11b}. The observed black hole-to-bulge mass ratio in NGC4342 is $6.9^{+3.8}_{-2.3}\%$, which significantly exceeds the typically observed $\sim$$0.2\%$. Given the uncertainty of the black hole mass measurement and the scatter in the $M_{\bullet}-M_{\rm{bulge}}$ relation, NGC4342 is a $\approx$$5.1\sigma$  outlier from the mean $M_{\bullet}-M_{\rm{bulge}}$ relation (for details see Section \ref{sec:ratio}). Although the galaxy is projected on the Virgo cluster, it belongs to the $W'$ cloud at a distance of $\sim$$23$ Mpc \citep{mei07,bogdan12}. Therefore we adopt a distance of $D = 23$ Mpc ($1\arcsec=111$ pc) for NGC4342. The Galactic column density towards NGC4342 is $N_H=1.6 \times 10^{20} \ \rm{cm^{-2}}$ \citep{dickey90}. 

To extend our study  we also investigate an elliptical galaxy, NGC4291 \citep{devaucouleurs91}. This galaxy  exhibits a black hole-to-bulge mass ratio of $1.9\%\pm0.6\%$, hence taking into account the black hole mass measurement uncertainty and the scatter in the   $M_{\bullet}-M_{\rm{bulge}}$ relation, the overly massive black hole corresponds to a $\approx$$3.4\sigma$  outlier from the mean $M_{\bullet}-M_{\rm{bulge}}$ relation (Section \ref{sec:ratio}). NGC4291 is located in a poor group with 11 member galaxies \citep{garcia93}. For the distance of NGC4291 we adopted $D = 26.2$ Mpc ($1\arcsec=126$ pc) \citep{tonry01}.  The Galactic column density towards NGC4342 is $N_H=3.0 \times 10^{20} \ \rm{cm^{-2}}$ \citep{dickey90}. 

This paper is structured as follows. In Section 2, we introduce the data that we have used and describe its reduction. In Section 3, we compute the black hole-to-bulge mass ratios for NGC4342 and NGC4291. The existence of massive dark matter halos around the galaxies is demonstrated in Section 4. Our results are discussed in Section 5, and we summarize in Section 6 

\begin{figure*}
 \begin{center}
       \includegraphics[width=8.5cm]{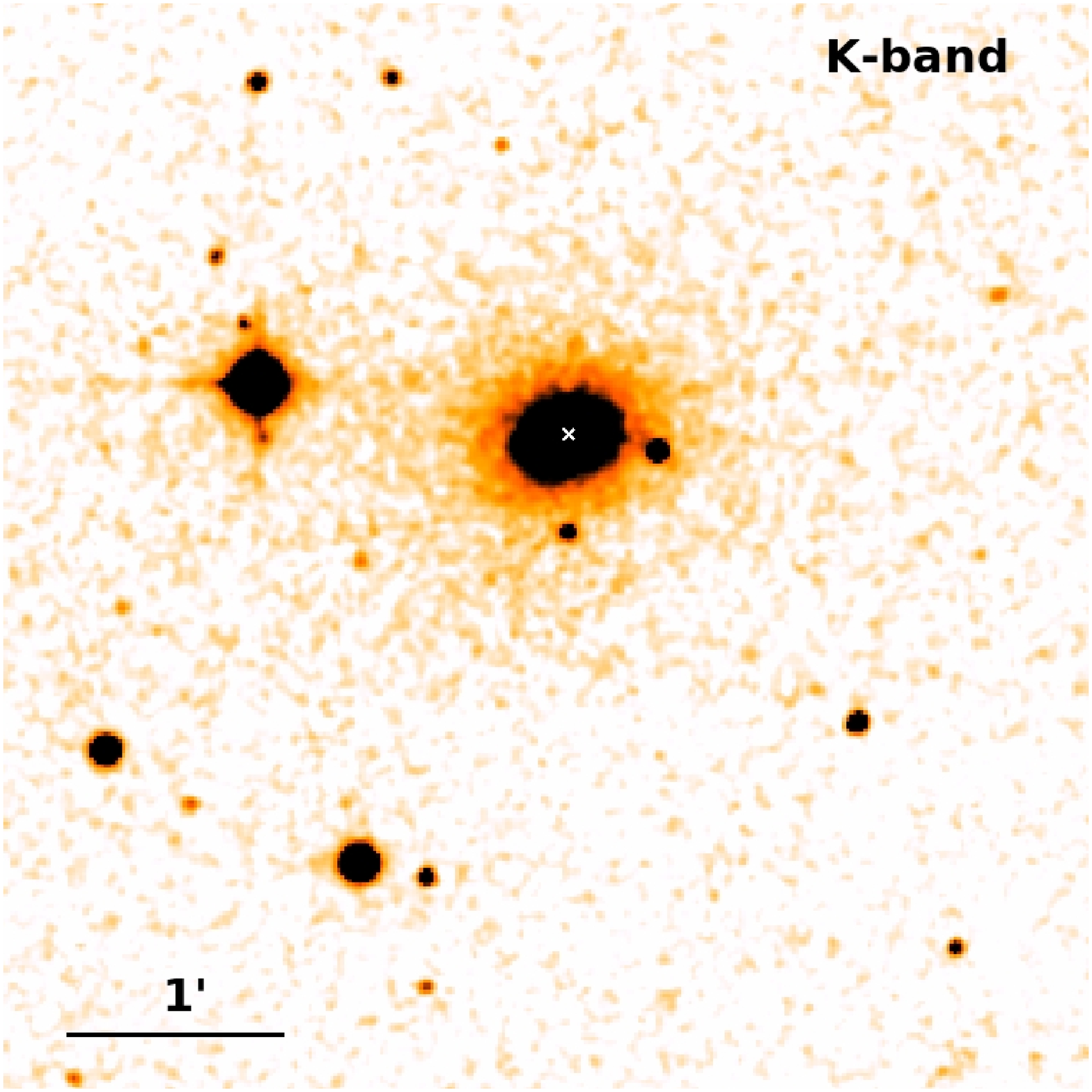}
\hspace{0.2cm}
      \includegraphics[width=8.5cm]{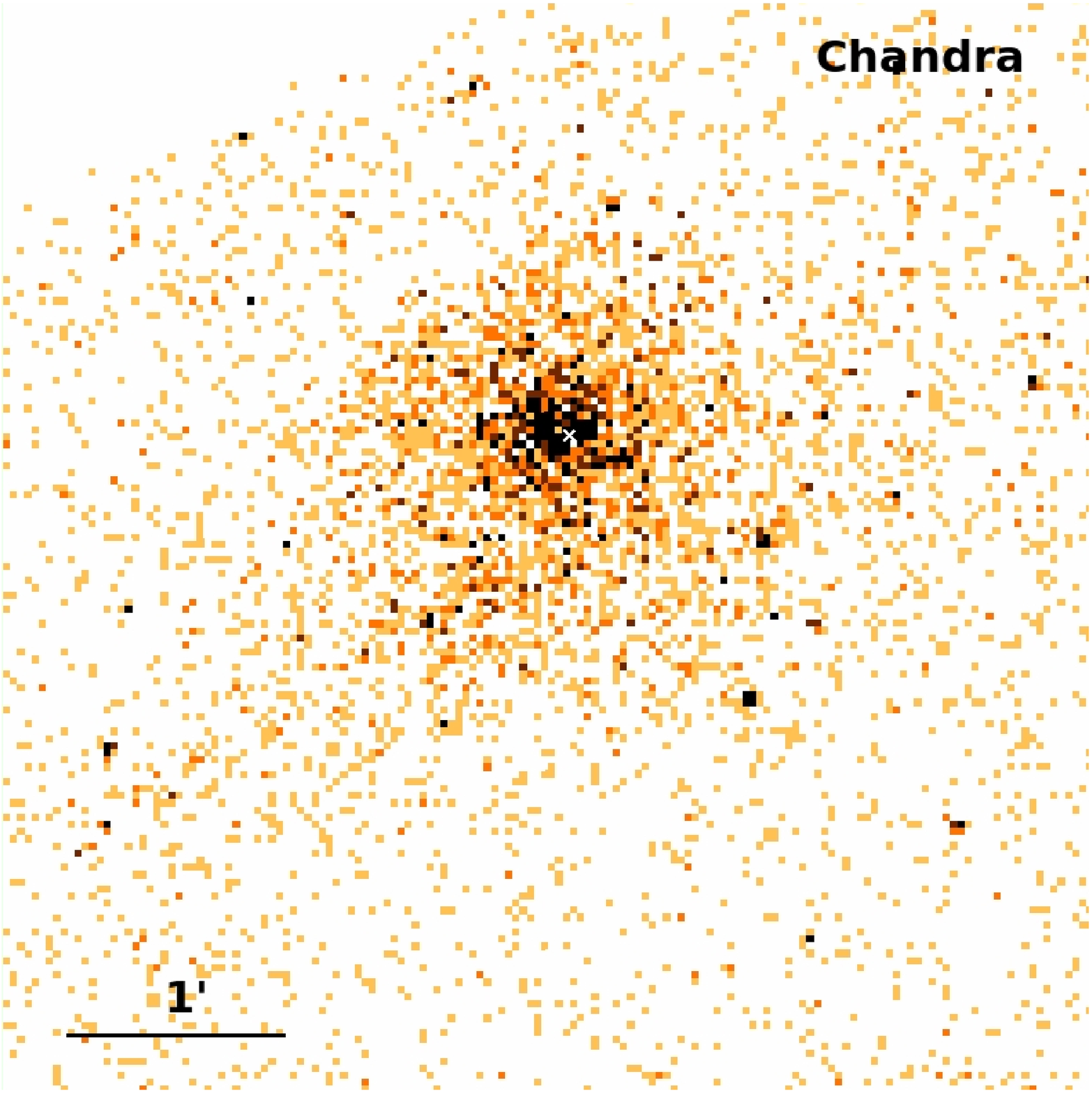}
       \caption{\textit{Left:} 2MASS K-band image of a $5\arcmin \times 5\arcmin$   (or $37.8 \times 37.8 $ kpc)  region around NGC4291. The center of the galaxy is marked with a white cross. The bright object to the east of the galaxy is a foreground star.  \textit{Right:} $0.5-2$ keV band \textit{Chandra} image of the same region. One pixel corresponds to $\approx$$2$ arcsec. The X-ray light has a much broader distribution than the K-band light. The diffuse X-ray emission appears to be more extended towards the south. North is up and east is left.}
     \label{fig:ngc4291}
  \end{center}
\end{figure*}

\section{Data reduction}
\subsection{\textit{Chandra}}
NGC4342 was observed by the \textit{Chandra X-ray Observatory} in two observations (ObsIDs: 4687, 12955) for a total of $112.5$ ks (Figure \ref{fig:ngc4342} right panel), whereas NGC4291 was observed in one pointing (ObsID: 11778) for $30.1$ ks  (Figure \ref{fig:ngc4291} right panel). The pointing for both galaxies is on the ACIS-S3 CCD, however for the analysis we used all available CCDs, namely ACIS-I2/I3 and ACIS-S1/S2/S3. 

Major steps of the data analysis were performed as described in \citet{vikhlinin05}. In particular, we reprocessed the data set and applied the latest calibration corrections to the detected X-ray photons. We produced light curves for each observations to detect and remove high background periods \citep{markevitch03}. After cleaning the flare contaminated time periods the clean exposure times were  $78.8$ ks and $26.2$ ks  for NGC4342 and NGC4291, respectively. 

The diffuse emission fills the entire field-of-view of the ACIS-S3 CCD for both galaxies, therefore we determined the background level using ``blank-sky'' observations\footnote{http://cxc.harvard.edu/contrib/maxim/acisbg/}. Although the level of instrumental background varies slightly with time, its spectrum remains unchanged \citep{hickox06}. To account for these variations  we renormalized the background level using the $10-12$ keV band count rate ratios, in which energy range the effective area of \textit{Chandra} is virtually negligible. Since we aim to study the truly diffuse emission from NGC4342 and NGC4291, emission from  bright resolved sources need to be subtracted. Therefore, we ran the \textsc{CIAO wavdetect} tool in the $0.5-8$ keV energy range to generate a source list. To obtain large source cells and exclude $\gtrsim$$98\%$ of the source counts, we changed several parameters from their default values, which are discussed in \citet{bogdan12}. In total we removed 126 and 43 sources from NGC4342 and NGC4291, respectively. The resulting source cells were used to mask out the location of bright point sources for further analysis of the diffuse emission.

\subsection{Two-Micron All Sky Survey}
\label{sec:2mass}
To study the stellar mass distribution we rely on the K-band images of the Two-Micron All Sky Survey (2MASS) Large Galaxy Atlas (LGA) \citep{jarrett03}, which is known to be an excellent stellar mass tracer. The publicly available K-band images of NGC4342 (Figure \ref{fig:ngc4342} left panel) and NGC4291  (Figure \ref{fig:ngc4291} left panel) are not background subtracted, therefore to estimate and subtract the background level we use several nearby regions off the galaxy. 

To convert the K-band luminosity to stellar mass, we compute the K-band mass-to-light ratios based on the $ B -V $ color indices and results of galaxy evolution modeling following \citet{bell03}:
$$\log (M_{\star}/L_K) = -0.206 + (0.135 \times (B-V)).$$
For NGC4342 we use its luminosity weighted $ B -V = 0.86$ color index (Mihos et al. in preparation) that results in $M_{\star}/L_K=0.81$, whereas in case of NGC4291 we adopt $ B -V = 0.93$ \citep{devaucouleurs91} yielding  $M_{\star}/L_K=0.83$. The intrinsic scatter of the \citet{bell03} relation is $\sim$$0.1$ dex, which determines the accuracy of the obtained mass-to-light ratios. The main sources of systematic uncertainties are the effects of dust and starbursts, which may generate $\sim$$25\%$ uncertainty. The origin and significance of uncertainties is discussed in full particulars in \citet{bell03}.

\subsection{Deep optical image}
NGC4342 and its surroundings were observed as part of a deep, wide-field optical survey of the Virgo cluster by Mihos et al. (in preparation). The field surrounding M49 and NGC4365 was imaged in Spring 2011 using the Burrell Schmidt telescope at Kitt Peak National Observatory. The imaging data consists of $73$ $1200$ s images, each with a $1.65\degr\times1.65\degr$ field-of-view, dithered randomly by roughly half a degree between exposures. The individual images are then flattened using a master dark sky flat created from a median combine of 71 offset sky images taken at the same hour angle and declination as the Virgo data. The extended wings of bright stars are modeled and subtracted from each image \citep{slater09}, after which the images are sky subtracted by fitting a plane to the sky after masking out discrete sources. Finally, the images are scaled to a common photometric zeropoint and median combined together to make the master mosaic with a native resolution of $1.45$ arcsec/pixel. 

To bring out the faintest features, we mask individual sources (stars and galaxies, both bright and faint) in the master mosaic, then rebin the image to $13$ arcsec/pixel using a spatial median filter. In this paper we use a $1$ square degree cutout of this rebinned deep mosaic, centered on NGC 4365. Because NGC 4365 lies near the edge of the Mihos et al. survey field, the total exposure time drops sharply across this portion of the mosaic, from 10 hours near NGC 4365 to 3 hours in the south-west corner of the field. More details on the imaging strategy and data reduction techniques for the Virgo imaging survey can be found in \citet{mihos05} and \citet{rudick10}.

\section{Black hole-to-bulge mass ratios}
\label{sec:ratio}
\subsection{$M_{\bullet}/M_{\rm{bulge}} $ of NGC4342}
Stellar dynamical models of NGC4342 demonstrate that it harbors a supermassive black hole, whose mass (converted to $23$ Mpc distance) is $M_{\bullet}=4.6^{+2.6}_{-1.5} \times 10^{8} \ \rm{M_{\odot}}$ \citep{cretton99}. To estimate the stellar mass of the bulge  we use the decomposition of \citet{vika12}, who found that $\sim$$1/3$ of the total stellar light originates from the bulge. \citet{vika12} computed that the apparent K-band magnitude of the bulge is $10.3$ mag, which corresponds to a K-band luminosity of $L_K=8.3\times10^{9} \ \rm{L_{K,\odot}}$. To convert the luminosity to stellar mass we apply the K-band stellar-mass-to-light ratio of $M_{\star}/L_K=0.81$ (Section \ref{sec:2mass}), hence the bulge mass of NGC4342 is $ M_{\rm{bulge}} = 6.7 \times 10^{9} \ \rm{M_{\odot}} $. Thus, the black hole-to-bulge mass ratio of NGC4342 is $M_{\bullet}/M_{\rm{bulge}} = 6.9^{+3.8}_{-2.3}\%$.

\subsection{$M_{\bullet}/M_{\rm{bulge}} $ of NGC4291}
The supermassive black hole of NGC4291 has been studied recently by \citet{schulze11}, who used stellar dynamical modeling to compute the black hole mass. They included the effect of a dark matter halo in their models, which resulted in a mass larger than previously reported \citep{gebhardt03}. Namely, the mass of the nuclear black hole in NGC4291 (for a distance of $26.2$ Mpc) is $M_{\bullet}=(9.6 \pm 3.0) \times 10^{8} \ \rm{M_{\odot}}$ \citep{schulze11}.  The total apparent K-band magnitude of the galaxy is $8.417$ mag, which translates to a  K-band luminosity of $L_K=6.1\times10^{10} \ \rm{L_{K,\odot}}$. Based on the K-band mass-to-light ratio of $M_{\star}/L_K=0.83$  (Section \ref{sec:2mass}), the stellar mass of NGC4291 is $M_{\rm{bulge}}=5.1\times10^{10} \ \rm{M_{\odot}}$. Given these parameters the black hole-to-bulge mass ratio of NGC4291 is $M_{\bullet}/M_{\rm{bulge}}= 1.9\%\pm0.6\%$.

\begin{figure}
 \begin{center}
       \includegraphics[width=8.5cm]{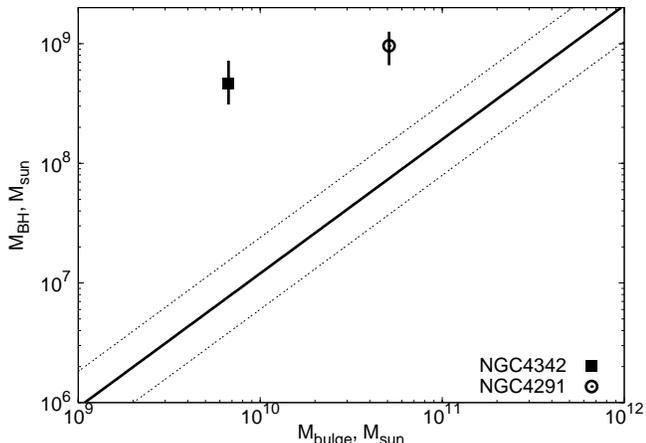}
       \caption{Black hole mass as a function of bulge mass. Thick solid line shows the mean $M_{\bullet} - M_{\rm{bulge}} $ relation from \citet{haring04}, whereas the thin dashed line represent the intrinsic scatter of the relation. Both NGC4342 and NGC4291 are highly significant outliers from the trend.}
     \label{fig:haring}
  \end{center}
\end{figure}

\begin{figure*}
 \begin{center}
       \includegraphics[width=8.5cm]{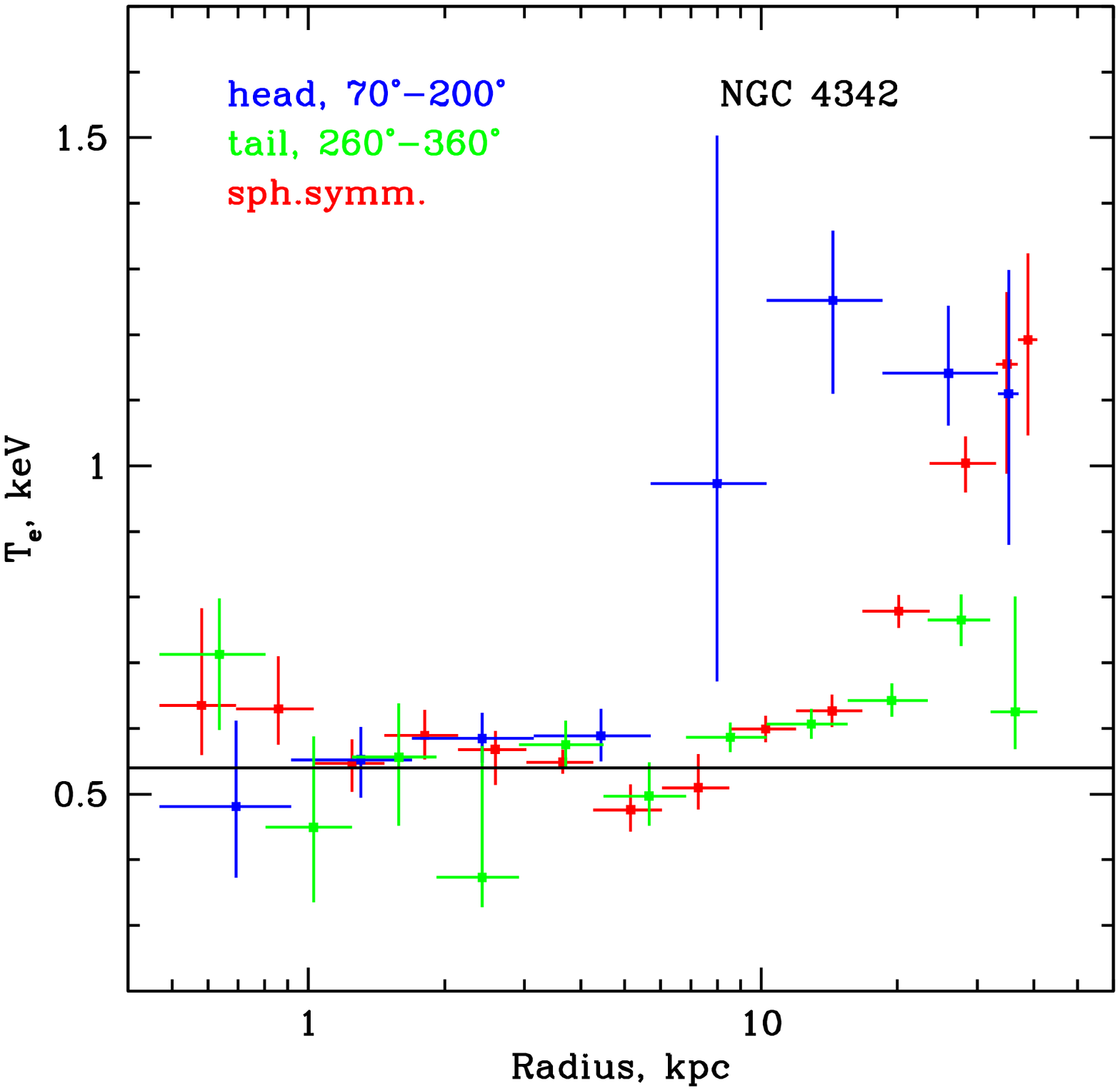}
\hspace{0.2cm}
     \includegraphics[width=8.5cm]{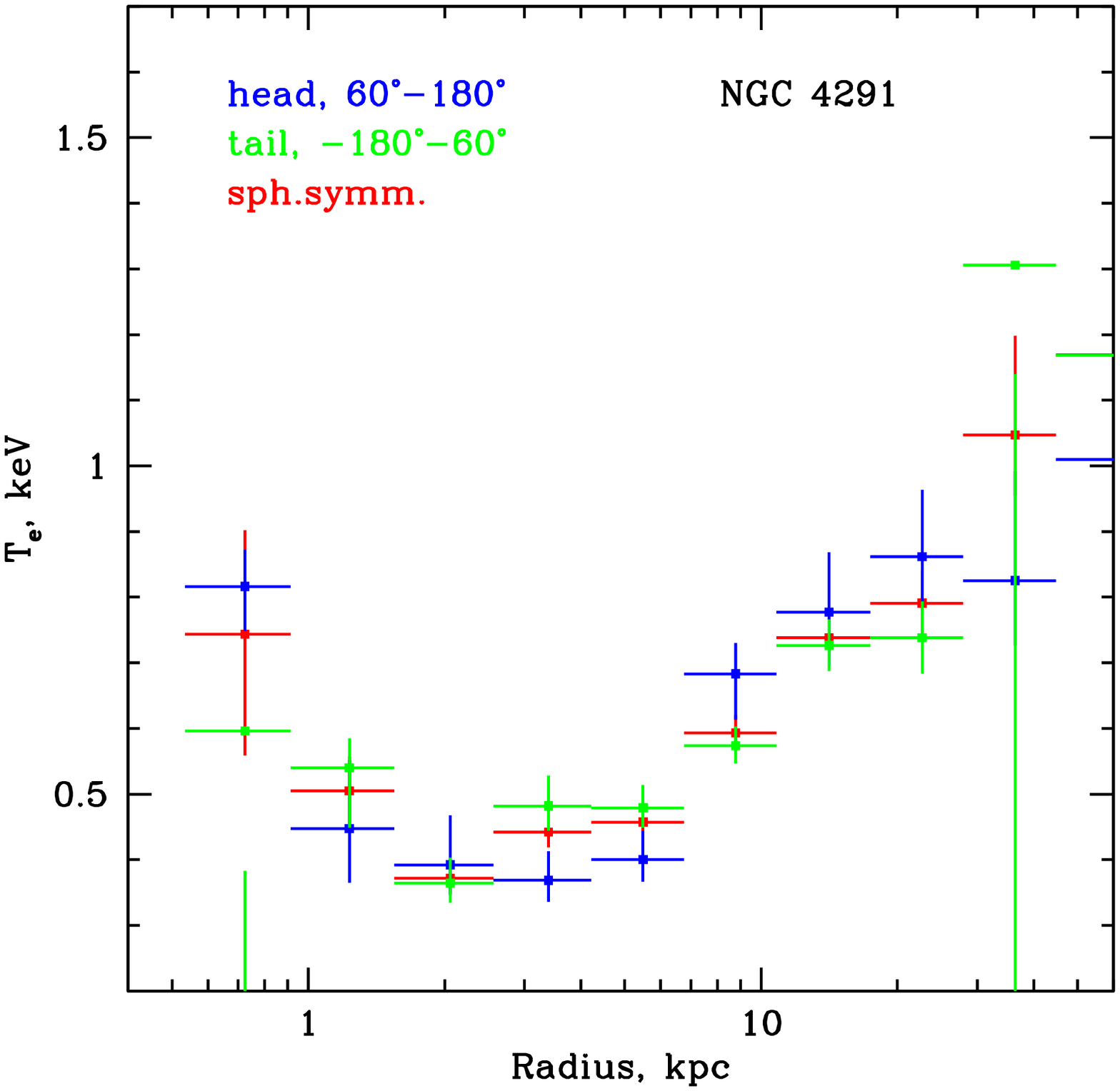}
       \caption{Projected temperature profiles for NGC4342 (left panel) and for NGC4291 (right). For both galaxies the profiles are depicted for three different sectors: (1) sectors towards the north(east) are shown with blue, (2) sectors towards the south(west) are green, and (3) circular annuli assuming spherical symmetry are red. The corresponding position angles for the sectors are labeled in the figures. The position angle $0\degr$ corresponds to west and increases counter-clockwise. For NGC4342 the solid horizontal line shows $kT=0.54$ keV, which was used to compute the gravitating mass profiles. To obtain the best-fit temperatures the abundances were fixed at $0.3$ and $0.1$ Solar for NGC4342 and NGC4291, respectively.}
     \label{fig:temp_profile}
  \end{center}
\end{figure*}

\subsection{Comparison with the mean $M_{\bullet}-M_{\rm{bulge}}$ relation}
\label{sec:ratio}
The observed black hole-to-bulge mass ratios in NGC4342 and NGC4291 are unusually large compared to other galaxies. To illustrate this point we show the mean  $M_{\bullet}-M_{\rm{bulge}}$ relation from \citet{haring04} in Figure \ref{fig:haring}, with the loci of NGC4342 and NGC4291 marked. Based on the \citet{haring04}  relation and the bulge masses of the two galaxies, the expected black hole masses are $7.7 \times 10^{6} \ \rm{M_{\odot}}$ and $7.4 \times 10^{7} \ \rm{M_{\odot}}$ in NGC4342 and NGC4291, respectively. Thus, the observed values are factors of $\approx$$60$ and $\approx$$13$ times larger than the predicted ones. From the intrinsic scatter of the relation ($0.30$ dex) and the uncertainty of the black hole mass measurements ($0.18$ dex and $0.12$ dex),  we conclude that NGC4342 and NGC4291 are $\approx$$5.1\sigma$  and $\approx$$3.4\sigma$ outliers, respectively.

Reversing the problem, we also compute the bulge masses, in which the supermassive black holes of NGC4342 and NGC4291 would be typical. According to the \citet{haring04} relation and the observed black hole masses, we find that the black holes of NGC4342 and NGC4291 would be expected in bulges with  $2.6 \times 10^{11} \ \rm{M_{\odot}}$ and $5.0 \times 10^{11} \ \rm{M_{\odot}}$ mass, respectively. These bulge masses exceed by factors of $\approx$$39$ and $\approx$$10$ the observed values in NGC4342 and NGC4291, respectively.

\section{Tidal stripping as a possible origin of the high black hole-to-bulge mass ratio}
One possibility to explain the unusually large black hole-to-bulge mass ratios observed in NGC4342 and NGC4291 is that most ($\gtrsim$$90\%$) of their stars were tidally stripped. However, the tidal stripping process would remove not only the stellar content of galaxies, but also the more loosely bound dark matter halos. Therefore, if $\sim$$90\%$ of the stars were stripped from the galaxies, no significant dark matter halos should be observed around them. Thus, to test the stripping scenario it is critial to determine whether NGC4342 and NGC4291 host extended dark matter halos.

In Section \ref{sec:halo}  we use \textit{Chandra} X-ray observations of the hot gas content of NGC4342 and NGC4291, to show that they reside in massive dark matter halos, thereby excluding the stripping scenario. In  Section \ref{sec:tail} we present a deep optical image of the surroundings of NGC4342, which -- independently from X-ray observations -- excludes the possibility that majority of the NGC4342 stellar population was tidally stripped.

\subsection{Dark matter halos}
\label{sec:halo}
\subsubsection{NGC4342}
The $0.5 - 2$ keV band X-ray image of NGC4342 reveals a diffuse hot gas component associated with the galaxy, which exhibits a significantly broader distribution than the stellar light (Figure \ref{fig:ngc4342}). To compute the gravitating mass profile of NGC4342, we assume that the hot gas is in hydrostatic equilibrium \citep{mathews78,forman85,humphrey06} and use the following equation:
$$  M_{\rm{tot}} (<r) = - \frac{kT_{\rm{gas}}(r) r }{G \mu m_{\rm{p}}}  \Bigg( \frac{\partial \ln n_{\rm{e}}}{\partial \ln r} + \frac{\partial \ln T_{\rm{gas}}}{\partial \ln r} \Bigg) ,   $$
where $T_{\rm{gas}}$ and $n_{\rm{e}}$ are deprojected temperature and density, respectively.  To determine the projected  temperature and density profiles of the hot gas, we describe the soft-band emission with an optically-thin thermal plasma emission model (\textsc{APEC} model in \textsc{Xspec}). To obtain deprojected profiles, we use the technique described by \citet{churazov03}. Namely, we model the observed spectra as the linear combination of spectra in spherical shells plus the contribution of the outer layer. We assume that emissivity in the outer shell declines as a power law with radius at all energies. The matrix that describes the projection of the shells into annuli is inverted and the deprojected spectra are calculated by applying the inverted matrix to the observed spectra.

Due the head-tail distribution of the hot X-ray emitting gas \citep[Figure \ref{fig:ngc4342} right panel;][]{bogdan12}, the assumption of hydrostatic equilibrium is questionable at radii larger than $\sim$$5$ kpc. To account for these uncertainties we computed mass profiles in three different sectors: (1) towards the northeast, (2) towards the southwest, and (3) assuming spherical symmetry. The left panel of Figure \ref{fig:temp_profile} illustrates that within the central $10$ kpc region of NGC4342 the hot gas is approximately isothermal. Additionally, in \citet{bogdan12} we show that the abundance is also fairly uniform within this region. Therefore to compute the mass profiles, we fix the gas temperature at $kT=0.54$ keV, the abundance at $0.3$ Solar \citep{grevesse98}, and the column density at the Galactic value \citep{dickey90}. We stress that beyond $\sim$$10$ kpc the temperature of the hot gas is non-uniform, furthermore deviations from the hydrostatic equilibrium significantly temper the accuracy of the mass measurements. Therefore, we rely only on the central $10 $ kpc region to estimate the  gravitating mass of NGC4342. We note that within this region the total number of counts in the $0.5-2$ keV band originating from the diffuse emission is $4713$, which yields a signal-to-noise ratio of $\approx$$50.8$, allowing statistically accurate measurements of the mass profile.

\begin{figure*}
 \begin{center}
       \includegraphics[width=8.5cm]{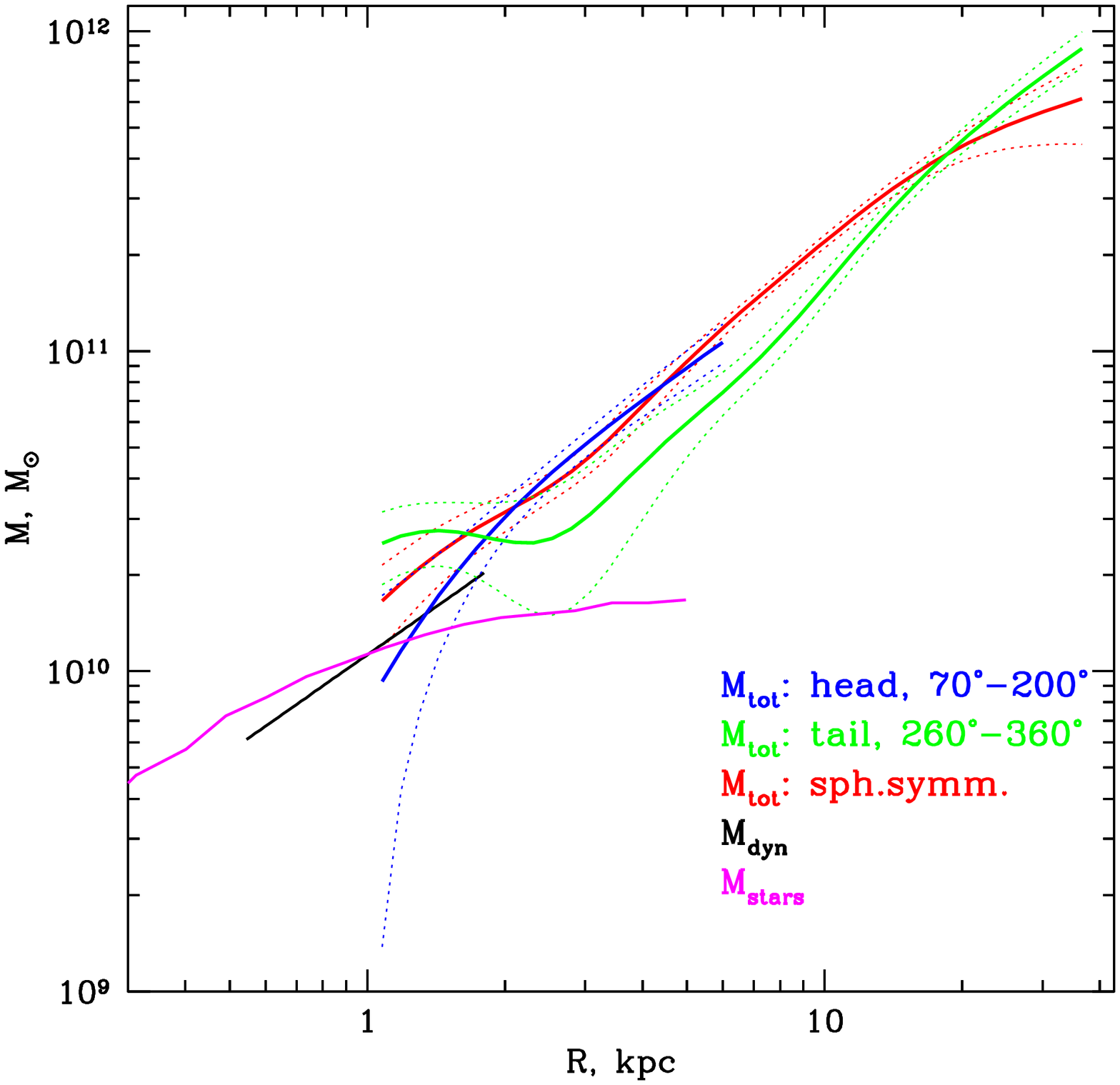}
\hspace{0.2cm}
     \includegraphics[width=8.5cm]{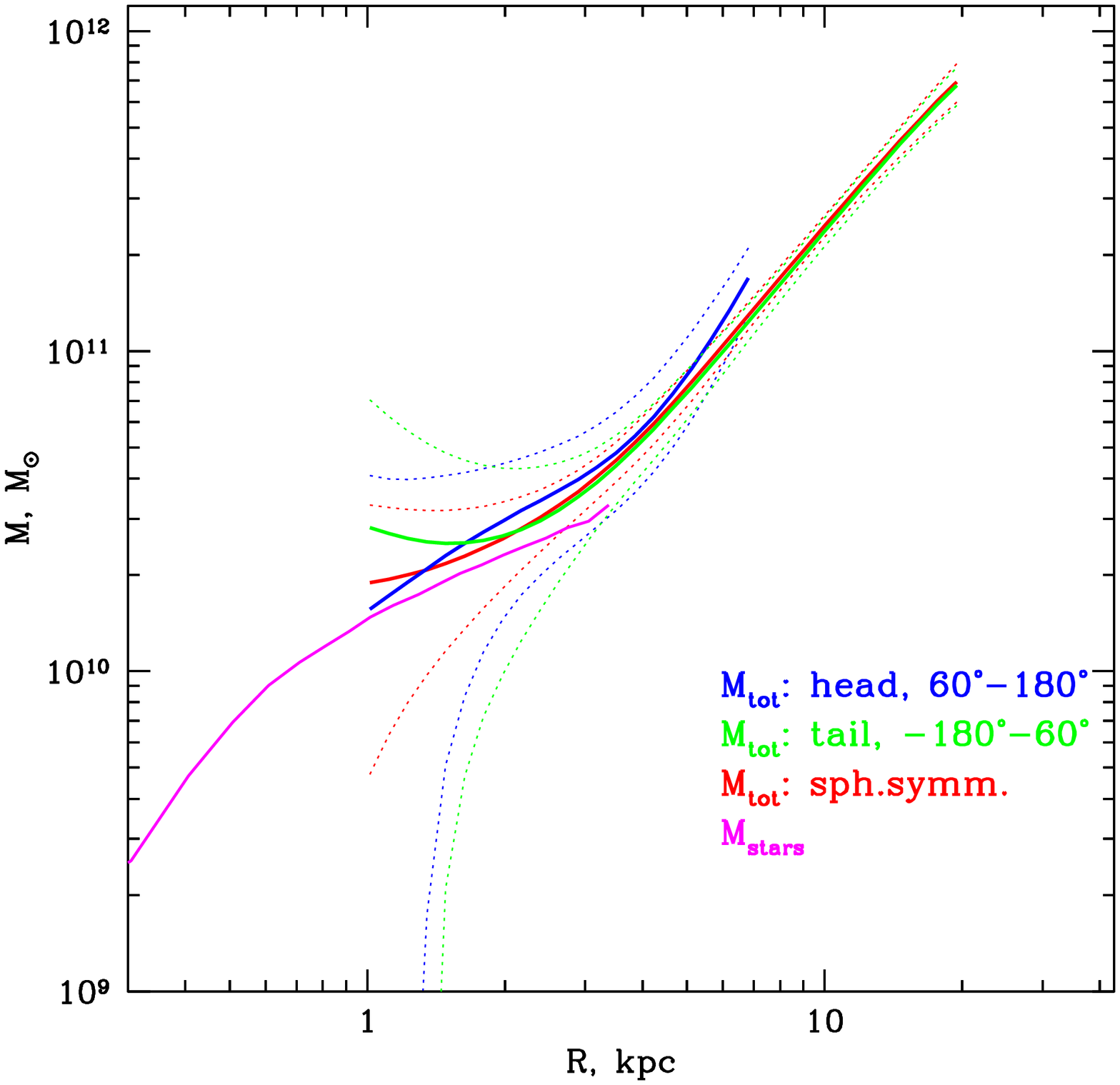}
       \caption{Gravitating mass profiles ($M_{\rm{tot}}$) of NGC4342 (left panel) and NGC4291 (right) based on \textit{Chandra} X-ray observations assuming that the hot gas is in hydrostatic equilibrium. We compute the gravitating mass in sectors towards the north(east) is blue, towards the south(west) of the galaxy is green, and assuming spherical symmetry is red. The solid lines represent the mean values, whereas the dotted lines indicate the lower and upper level of expected $1\sigma$ uncertainties, which were calculated using bootstrapping method. The pink solid curves ($M_{\rm{stars}}$) represent the stellar mass profiles obtained from the 2MASS K-band images. The black solid curve ($M_{\rm{dyn}}$) on the left panel shows an estimate of the total gravitating mass assuming flat circular speed of $220 \ \rm{km \ s^{-1}}$ adopted from \citet{cretton99}. Note that in the plotted regions the contribution of random motions is not significant.}
     \label{fig:mass}
  \end{center}
\end{figure*}

The mass profiles obtained for NGC4342 are shown in the left panel of Figure \ref{fig:mass}, where we also show the $1\sigma$ statistical uncertainties assuming that measurements in each radial bin are independent. In the same panel we also depict the stellar mass profile computed from the K-band luminosity using a mass-to-light ratio of $M_{\star}/L_K=0.81$, and the dynamical mass calculated assuming a circular speed of $220 \ \rm{km \ s^{-1}}$ \citep{cretton99}. Note that the circular speed was fixed in a certain region (approximately $5\arcsec < r < 12\arcsec $), where the contribution of random motions is not significant.  The left panel of Figure \ref{fig:mass} clearly demonstrates that beyond $\sim$$1$ kpc the stellar mass is significantly lower than the total gravitating mass. In particular, within the central $10$ kpc region the total gravitating mass is in the range $(1.4-2.3)\times10^{11} \ \rm{M_{\odot}}$, which exceeds the stellar mass by an order of magnitude, implying the existence of a significant dark matter halo around the galaxy.

Besides \textit{Chandra} observations, the metallicity of the stellar population of NGC4342 indirectly and independently indicates that it formed in a massive dark matter halo. In the commonly accepted picture, galaxies with massive dark matter halos are able to retain a significant fraction of the $\alpha-$elements produced by core collapse supernovae at early epochs of galaxy formation, whereas in galaxies with low mass halos, $\alpha-$elements are largely driven away in stellar winds \citep{tremonti04,gallazzi06}. At later epochs, Type Ia Supernovae enrich all galaxies with iron-peak elements, hence high $\alpha$-to-iron metallicity ratios (or the often used [Mg/Fe] abundance ratios -- \citealp{gonzalez93}) are indicative of massive dark matter halos. NGC4342 bears a high [Mg/Fe] $= 0.60\pm 0.04$ ratio \citep{bosch98}, which is typical for galaxies with dark matter halos of a few times $10^{12} \ \rm{M_{\odot}}$ \citep{vazdekis04,humphrey06}.

\subsubsection{NGC4291}
In Figure \ref{fig:ngc4291} we show the K-band and the $0.5-2$ keV band \textit{Chandra} images of NGC4291. The truly diffuse X-ray emission associated with NGC4291 originates from hot X-ray gas and extends well beyond the optical light. To compute the gravitating mass profiles we assume that the hot X-ray gas is in hydrostatic equilibrium and follow the procedure described for NGC4342. Since the X-ray emission shows asymmetries, which may indicate deviations from hydrostatic equilibrium, we also computed the gravitating mass profiles using three  different sectors: (1) towards the north, (2) towards the south, and (3) assuming spherical symmetry. The temperature of the X-ray gas shows variations within the central $10$ kpc region (Figure \ref{fig:temp_profile} right panel), therefore when computing the mass profiles, the temperature was left free to vary. The abundance of the hot gas is fairly uniform at $0.1$ Solar, therefore this parameter was fixed. Since the temperature cannot be determined accurately at radii larger than $\sim$$10$ kpc, and the distribution of hot gas cannot be constrained at these larger radii, we only rely on the central $10 $ kpc region of NGC4291 to compute the gravitating mass profiles. Within the $10$ kpc region we observed $3258$ counts in the $0.5-2$ keV band from the diffuse emission, hence the signal-to-noise ratio is high, $\approx$$49.7$. 

The resulting mass profiles with the $1\sigma $ uncertainties are shown in the right panel of Figure \ref{fig:mass}. In the figure  we also plot the stellar mass distribution obtained from the 2MASS K-band image using the mass-to-light ratio of $M_{\star}/L_K=0.83$. The plot illustrates that the stellar mass only plays an important role in the central $\sim$$2$ kpc, beyond which radius the dark matter halo becomes dominant. Within $10$ kpc the total mass is $(2.1-2.6)\times10^{11} \ \rm{M_{\odot}}$, which is a factor of $\sim$$5$ times higher than the stellar mass of the galaxy. We thus conclude that NGC4291 resides in a substantial dark matter halo.

\subsubsection{Excluding the tidal stripping scenario}
Although we cannot reliably characterize the dark matter halo of NGC4342 and NGC4291 at radii much larger than $\sim$$10$ kpc, \textit{Chandra} measurements  demonstrate that both galaxies host massive dark matter halos, which extend well beyond the stellar light. 

In general, the stellar populations of galaxies are significantly more tightly bound than the associated dark matter halos. The dominant fraction of the stellar population of galaxies is concentrated in the central $\sim$$5$ kpc region, whereas the dark matter halo may extend up to a few $100$ kpc \citep{humphrey06}. If $\sim$$90\%$ of the stars were lost from NGC4342 or NGC4291, they were removed from the central few kpc region. However, owing to its low specific binding energy the dark matter halo will be stripped first and faster than the stars \citep{libeskind11}. Thus, the detection of a significant dark matter halo is inconsistent with the tidal stripping scenario.

\begin{figure}
 \begin{center}
       \includegraphics[width=8.5cm]{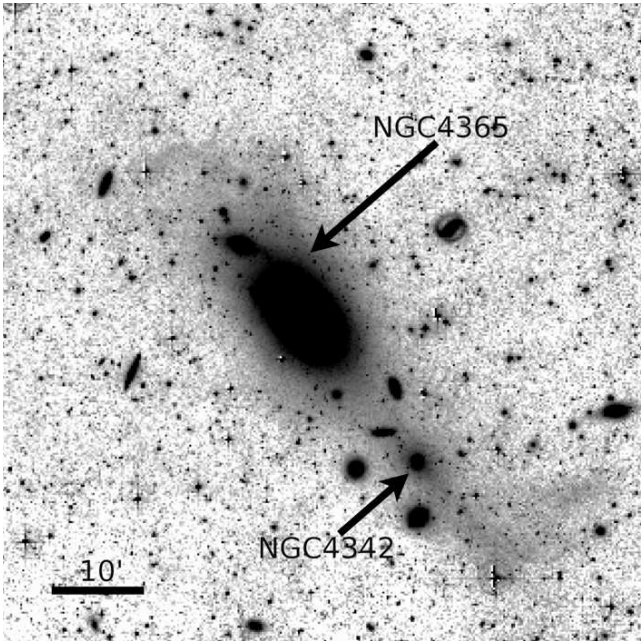}
       \caption{Portion of the deep B-band Virgo mosaic (Mihos et al. in prep.), showing a one square degree region around NGC4365. The mosaic has a native resolution of $1.45$ arcsec/pix, but this image has been rebinned using a median filter to $13$ arcsec/pix resolution to emphasize faint features. The limiting surface brightness in the image is $\mu_B \approx 29.5 \ \rm{mag/arcsec^2}$, and regions brighter than $\mu_B \approx 26 \ \rm{mag/arcsec^2}$ are rendered black. A prominent feature on this image is a broad, $\sim$$200$ kpc long tidal tail extending southwest from NGC4365 with a mean surface brightness of $\mu_B \approx 28 \ \rm{mag/arcsec^2}$. A fainter and smaller tidal loop can also be seen to the northeast of NGC4365.}
     \label{fig:tail}
  \end{center}
\end{figure}

\subsection{Tidal tail behind NGC4365}
\label{sec:tail}
The deep B-band image, shown in Figure \ref{fig:tail}, is also in conflict with the tidal stripping scenario. The image unveils an extended, faint tidal tail behind NGC4365. Interestingly, several galaxies, including NGC4342, appear to be embedded in this tidal tail. NGC4365 lies $0.33\degr$ ($130$ kpc in projection) northeast from NGC4342, and is presumably the dominant member of the $W'$ cloud \citep{mei07}. Although the exact origin of the tidal tail is uncertain, its presence and its distinct structure suggest that NGC4365 underwent a relatively recent ($\lesssim 2 \times 10^9$ years) interaction or merger with a companion galaxy.

If $\sim$$2 \times 10^{11} \ \rm{M_{\odot}} $ stars were stripped from NGC4342 during an interaction with NGC4365, a large fraction of the missing stars should be still observable around the galaxy since NGC4342 has an $\sim$$8$ Gyrs old stellar population \citep{bosch98}. However, the missing stars are unaccounted for, since the integrated B-band magnitude of the tidal tail behind NGC4365 (Figure \ref{fig:tail}) is $\sim$$13.5$ mag, which corresponds to $L_B\approx3.6\times10^{9} \ \rm{L_{\rm{B,\odot}}}$. We convert this luminosity to stellar mass using the B-band mass-to-light ratio of $M_{\star}/L_B=3.6$, which we estimate under the (extreme) assumption that the tail has been stripped from NGC 4342, and using the $B-V=0.86$ color index of NGC4342 and results of galaxy evolution modeling \citep{bell03}. Hence the total mass of the tidal tail behind NGC4365 is $1.3 \times 10^{10} \ \rm{M_{\odot}} $ – more than an order of magnitude less than the  ``missing''  stellar mass from NGC4342. Thus, the optical image, independently from \textit{Chandra} observations, disproves that $\sim$$90\%$ of the original stellar content of NGC4342 was stripped.

\section{Discussion}
The exclusion of the stripping scenario implies that dominant fraction of the stellar population has not been lost as theorized in the tidal stripping scenario, but in fact never formed, making NGC4342 and NGC4291 genuine outliers from the $M_{\bullet}-M_{\rm{bulge}}$ relation. This conclusion has important consequences for the formation and evolution of black holes and galaxies. 

The presence of unusually massive black holes in two low stellar mass galaxies indicates that the growth of the supermassive black holes and the bulges of NGC4342 and NGC4291 was not tightly coupled. In particular, our results  imply that the growth of the supermassive black hole can precede the major burst of star formation. Although this concept is in conflict with the standard theoretical paradigm of bulge-black hole co-evolution, in which bulges and black holes co-evolve, a number of  observational and theoretical studies also support our conclusions. In a recent study \citet{reines11} demonstrated the presence of a supermassive black hole in Henize 2-10 (a starburst galaxy without a substantial bulge) suggesting that supermassive black holes can form before their bulges. The evolution of the $M_{\bullet}-M_{\rm{bulge}}$ scaling relation at high redshifts also hint that black holes may grow before their host bulges \citep[e.g.][]{walter04,jahnke09,merloni10}. Additionally, numerical simulations of quasar hosts indicate that, at high redshifts ($5.5\lesssim z \lesssim6.5$), black holes are more massive for their stellar hosts than expected from the relations observed in the local Universe \citep{khandai11}. In good agreement with this, \citet{petri12} modeled the early formation of supermassive black holes and obtained similar results, namely that at high redshifts the $M_{\bullet}-M_{\rm{bulge}}$ relation was steeper than presently observed, hinting that  black holes may grow faster than their host bulges. 

A major result of this paper is that both NGC4342 and NGC4291 reside in massive dark matter halos. In fact, both the black hole masses and the observed dark matter halos are typical of galaxies having stellar masses that are $\sim$$10-40$ times greater, hence the characteristics of the dark matter halos are consistent with those expected for the black holes. Therefore the only truly  anomalous property of NGC4342 and NGC4291 are their low stellar masses. Since the black hole mass  correlates well with the halo mass, it suggests that dark matter halos may play a fundamental role in governing the black hole growth. Indeed, the tight relation between the black hole mass and the dark matter halo mass has already been recognized by \citet{ferrarese02}. Although \citet{kormendy11} hinted based on a sample of low-mass galaxies that supermassive black holes do not correlate with dark matter halos, \citet{volonteri11} re-examined their results and concluded that properties of dark matter halos do play an important role in the  assembly of supermassive black holes.  The arguments of \citet{volonteri11} are further supported by numerical studies, which demonstrate that the $M_{\bullet}-M_{\rm{bulge}}$ relation can be produced through the central limit theorem \citep{peng07,jahnke11}, hence the existence of the scaling relation does not necessarily invoke any causal process between the black hole and bulge evolution. Moreover, numerical simulations of \citet{booth10} pointed out that the primary parameter in determining the black hole mass is the mass of the dark matter halo. 

Although NGC4342 and NGC4291 are outliers from the $M_{\bullet}-M_{\rm{bulge}}$ relation, both galaxies fit  the $M_{\bullet}-\sigma_{\rm{c}}$ relation rather well. The central velocity dispersions of NGC4342 and NGC4291 are $252.1\pm 8.4 \ \rm{km \ s^{-1}}$ and   $285.3 \pm 5.7 \ \rm{km \ s^{-1}}$, respectively \citep[][HyperLeda\footnote{http://leda.univ-lyon1.fr/}]{prugniel96}. Based on the velocity dispersions and the mean $M_{\bullet}-\sigma_{\rm{c}}$ relation \citep{gultekin09}, the expected black hole masses for NGC4342 and NGC4291 are $4.2 \times 10^{8} \ \rm{M_{\odot}} $ and  $6.9 \times 10^{8} \ \rm{M_{\odot}} $, which are in good agreement with the observed values. This demonstrates that the velocity dispersion is a better proxy for the black hole mass than the bulge mass \citep[see also][]{rusli11}. Moreover, the tight relation between the black hole mass and the velocity dispersion is also in good agreement with the self-regulated growth of supermassive black holes \citep{silk98,fabian99,king03,dimatteo05}. 

The observed physical properties ($M_{\bullet}/M_{\rm{bulge}}$ and $M_{\rm{\star}}/M_{\rm{total}}$ ratios) of NGC4342 classify it as one of the most extreme outliers in the local Universe. Therefore, it is tempting to examine whether theoretical models can reproduce the observed characteristics of NGC4342. We study the galaxy catalog constructed on top of the Millennium simulation \citep{guo10} and search for simulated counterparts of NGC4342. In particular, we searched for galaxies with $M_{\bullet}/M_{\rm{bulge}}>2.3\%$ ($2\sigma$ lower than observed for NGC4342) and with bulge mass in the range of $2\times10^9 \ \rm{M_{\odot}}<M_{\rm{bulge}} < 1\times10^{10} \ \rm{M_{\odot}}$. Several galaxies fulfilling these criteria are  tidally stripped systems, hence they do not host an extended dark matter halo. Among those galaxies with a notable dark matter halo, the highest observed black hole-to-bulge mass ratio is $M_{\bullet}/M_{\rm{bulge}}\lesssim3\%$, significantly lower than observed in NGC4342. Moreover, the simulated counterparts also host black holes with $M_{\bullet}\lesssim10^{8} \ \rm{M_{\odot}}$, and large fraction of the galaxies are strongly disk dominated. Thus, no galaxy in the Millennium simulation can reproduce the observed properties of NGC4342, thereby indicating that the evolution of NGC4342 is not fully explained by current theoretical models.

Although the presently observed $M_{\bullet}/M_{\rm{bulge}}$ ratios of NGC4342 and NGC4291 significantly exceed those observed in other bulges, it is possible that the evolution of these galaxies will eventually result more massive bulges, hence normal $M_{\bullet}/M_{\rm{bulge}}$ ratios. Very low level star formation may be present in the central regions of NGC4342 and NGC4291 \citep{rafferty08}, which process could increase the bulge mass. However, low level star-formation ($\ll1 \ \rm{M_{\odot} \ yr^{-1}}$) can produce the missing stellar mass, $\sim$$(2.5-4.4)\times10^{11} \ \rm{M_{\odot}}$, only in multiples of the Hubble time, moreover none of these galaxies host the required amount of either cold or hot gas to produce the missing stellar mass. Therefore, galaxy mergers may play a more important role in increasing the bulge mass of NGC4342 and NGC4291. However, the occurrence time scale of the sufficient number of (major-)mergers may be comparable or even longer than the Hubble time due to the relatively low merger rate at the present epoch \citep[e.g.][]{bower06,delucia07}.

Finally, based on the  conclusions above we describe a potential evolutionary scenario, which is consistent with and could result in the observed physical properties of NGC4342 and NGC4291.  The principal quantity determining the fate of the gas in the halo is the angular momentum. Indeed, large angular momentum promotes formation of the disk and more efficient star formation, but suppresses direct flow toward the central black hole. By contrast accretion onto a black hole and the coupling of the black hole feedback to the gas works best if the angular momentum of the gas is small \citep{churazov05}. Thus, it is plausible that NGC4342 and NGC4291 were formed in a sequence of accretion events, which brought mostly low angular momentum gas to the system. For details on the accretion of low angular momentum gas at the early-epochs of galaxy formation ($z\sim6$) refer to   \citet{dubois11}. The accretion of low angular momentum gas resulted in enhanced black hole growth, while the star formation without strong rotational support mostly happened in a relatively compact central region. At intermediate redshifts ($z\sim2-4$) the accretion of gas with higher angular momentum \citep{dubois11} could play a role in building the stellar body of NGC4342 and NGC4291.  Once the black hole was large enough, it prevented further gas cooling  \citep[e.g.][] {silk98,churazov01,dimatteo05,hopkins06,mcnamara07}, switching to the low accretion rate mode with very little star formation \citep{churazov05}. In this picture, the virial temperature of the gas/halo is probably the most important parameter governing the black hole growth, hence it is not surprising that the final mass of the black hole ``knows'' about the characteristics of the halo. Therefore despite the significant deviations from the black hole-to-bulge mass relation, NGC4342 and NGC4291 could still obey other relations, which are linked to their dark matter halo.

\section{Summary}
We studied two early-type galaxies, NGC4342 and NGC4291, with black hole-to-bulge mass ratios of $\approx$$6.9\%$ and $\approx$$1.9\%$, which are among the highest observed values in the local Universe. Our results can be summarized as follows: \\

1. Assuming that the hot X-ray emitting gas content of the galaxies is in hydrostatic equilibrium we computed gravitating mass profiles. The total mass within $10$ kpc is $(1.4-2.3)\times10^{11} \ \rm{M_{\odot}}$ and $(2.1-2.6)\times10^{11} \ \rm{M_{\odot}}$ for NGC4342 and NGC4291, respectively. These values significantly exceed the stellar mass, implying the presence of significant dark matter halos around the galaxies.

2. The detection of extended dark matter halos implies that tidal stripping, in which $\gtrsim$$90\%$ of the stellar mass was lost, cannot explain the observed high black hole-to-bulge mass ratios. In case of NGC4342 a deep optical image of its surroundings further supports this conclusion. 

3. Since tidal stripping can be excluded as the origin of the high back hole-to-bulge mass ratios, we conclude NGC4342 and NGC4291 were formed as low stellar mass systems and that the black hole and bulge did not grow in tandem. In particular, our results suggest that the black holes suppressed star formation and  the black hole possibly grew much faster  than the host bulges. 

4. Since the characteristics of the dark matter halos are consistent with those expected from the black hole, we concluded that the dark matter halos can play a fundamental role in governing the black hole growth. The black hole masses of NGC4342 and NGC4291 tightly correlate with the velocity dispersions, which indicates the self-regulated growth of the black holes. 

5. We described a potential evolutionary scenario that can produce the observed characteristics of NGC4342 and NGC4291. In this picture, due to the accretion of low angular momentum gas, the black hole grows more rapidly than the stellar population. Once the black hole reaches a critical mass, it prevents further gas cooling and star-formation. 

\bigskip
\begin{small}
\noindent
\textit{Acknowledgements.}
\'AB and WRF are grateful to Debora Sijacki, Mark Vogelsberger, Frank van den Bosch, Kevin Schawinski, and Charlie Conroy for helpful discussions and to Hans B\"ohringer for his valuable contribution to the \textit{Chandra} proposal. This research has made use of \textit{Chandra}  data provided by the Chandra X-ray Center. The publication makes use of software provided by the Chandra X-ray Center (CXC) in the application package CIAO. This publication makes use of data products from the Two Micron All Sky Survey, which is a joint project of the University of Massachusetts and the Infrared Processing and Analysis Center/California Institute of Technology, funded by the National Aeronautics and Space Administration and the National Science Foundation. \'AB acknowledges support provided by NASA through Einstein Postdoctoral Fellowship grant number PF1-120081 awarded by the Chandra X-ray Center, which is operated by the Smithsonian Astrophysical Observatory for NASA under contract NAS8-03060. WF and CJ acknowledge support from the Smithsonian Institution. GQ acknowledges support from the National basic research program of China (973 program under Grant No. 2009CB24901), the Young Researcher Grant of National Astronomical Observatories, CAS, the NSFC grants program (No. 11143005), and the Partner Group program of the Max Planck Society. JCM appreciates support from the NSF through grants AST-0607526, AST-0707793, and AST-1108964.
\end{small}

\end{document}